\documentclass[aps,groupedaddress,amsmath,showpacs]{revtex4}
\usepackage{bm}
\usepackage{amssymb}
\usepackage[draft]{graphicx}

\usepackage{bm}
\usepackage{amsmath}
 \usepackage{color}

\newcommand{\beq}{\begin{equation}}
\newcommand{\eeq}{\end{equation}}
\newcommand{\beqa}{\begin{eqnarray}}
\newcommand{\eeqa}{\end{eqnarray}}
\newcommand{\nn}{\nonumber\\}

\begin{document}

\title{Comment on `A general integral identity'}
\author{Andr\'es Santos}
\email{andres@unex.es}
\homepage{http://www.unex.es/eweb/fisteor/andres/}
\affiliation{Departamento de F\'{\i}sica, Universidad de
Extremadura, E-06071 Badajoz, Spain}

\date{\today}
\begin{abstract}
A simple heuristic proof of an integral identity recently derived (Glasser ML 2011  \emph{J. Phys. A: Math. Theor.} \textbf{44} 225202) is presented.
\end{abstract}
 \maketitle

In a recent paper \cite{G11}, Glasser has derived the integral identity
\beq
I(r)\equiv\int _0^{\pi /2}{\rm d}\theta \int _0^{\pi /2}{\rm d}\phi \sin \theta F(r\sin \theta \sin \phi  )=\frac{\pi }{2}\int
   _0^1{\rm d}t F(rt),
\label{1}
\eeq
where $F(u)$ is an arbitrary function, by following a relatively sophisticated method. 
At the end of his paper, Glasser states that `The results derived here also follow from the invariance of an integral over the surface
of an $n$-sphere under a permutation of the angular hyperspherical co-ordinates' and cites a private communication from the eminent physicist F J Dyson. 
The aim of this Comment is to provide a simple proof of Eq.\ \eqref{1} based on  symmetry arguments, thus fleshing out the details of the previous brief quotation. 

First, we make the change of variables $\theta\to\pi-\theta$ to obtain
\beqa
\label{2}
I(r)&=&\int_{\pi/2}^{\pi}{\rm d}\theta \int _0^{\pi /2}{\rm d}\phi \sin \theta F(r\sin \theta \sin \phi  )\nn
&=&\frac{1}{2}\int_{0}^{\pi}{\rm d}\theta \int _0^{\pi /2}{\rm d}\phi \sin \theta F(r\sin \theta \sin \phi  ).
\eeqa
Next, we perform the analogous change $\phi\to\pi-\phi$. Thus,
\beq
I(r)
=\frac{1}{4}\int_{0}^{\pi}{\rm d}\theta \int _0^{\pi}{\rm d}\phi \sin \theta F(r\sin \theta \sin \phi  ).
\label{3}
\eeq

Now, let us assume that $r>0$ (the case $r<0$ can be treated separately in a similar way). In that case, the arguments of the function $F$ in Eq.\ \eqref{1} are positive. This implies that, when proving Eq.\ \eqref{1}, only $F(u)$ for $u>0$ matters and thus we are free to extend $F(u)$ for $u<0$. If we extend it as an even function, i.e., $F(-u)=F(u)$, and make the change $\phi\to \phi+\pi$ in Eq.\ \eqref{3}, the result is
\beqa
I(r)&=&\frac{1}{4}\int_{0}^{\pi}{\rm d}\theta \int _{\pi}^{2\pi}{\rm d}\phi \sin \theta F(r\sin \theta \sin \phi  )\nn
&=&
\frac{1}{8}\int_{0}^{\pi}{\rm d}\theta \int _{0}^{2\pi}{\rm d}\phi \sin \theta F(r\sin \theta \sin \phi  ).
\label{4}
\eeqa
This can be recognized as an integration over all the spatial directions of the function $F$ evaluated at $y=r\sin \theta \sin \phi$ in (three-dimensional) spherical coordinates, i.e.,
\beq
I(r)=\frac{1}{8}\int {\rm d}\Omega F(y),
\label{5}
\eeq
where ${\rm d}\Omega={\rm d}\theta {\rm d}\phi \sin \theta $ is the elementary solid angle.

Thus far, only formal manipulations in the integrals defining $I(r)$ have been made. To conclude the proof of Eq.\ \eqref{1} we simply take into account that, by \emph{symmetry}, $I(r)$ must be independent of the choice of Cartesian axes, so that we can freely replace $F(y)$ by $F(x)$ or $F(z)$ inside the integral. In particular, the latter choice yields
\beqa
I(r)&=& \frac{1}{8}\int {\rm d}\Omega F(r\cos\theta)\nn
&=&\frac{2\pi}{8} \int_{0}^{\pi}{\rm d}\theta  \sin \theta F(r\cos\theta)\nn
&=&\frac{\pi}{4} \int_{-1}^{1}{\rm d}t   F(rt),
\label{6}
\eeqa
where the change $\theta\to t=\cos\theta$ has been made in the last step. Recalling the extension $F(-u)=F(u)$, Eq.\ \eqref{1} is finally obtained.

Support from the Ministerio de  Ciencia e Innovaci\'on (Spain) through Grant No.\ {FIS2010-16587} and the Junta de Extremadura (Spain) through Grant No.\ GR10158, partially financed by FEDER funds, is gratefully acknowledged.

\end{document}